\documentclass[fleqn]{2023SCGE}
\usepackage{graphicx}
\usepackage{bm}
\usepackage{lipsum}
\usepackage{amsfonts}
\usepackage{mathrsfs}
\usepackage{amsmath,amssymb,latexsym,bbm}
\usepackage{braket}
\usepackage{hepunits}
\usepackage[svgnames]{xcolor}
\usepackage{pifont}

\newcommand{\nn}{\nonumber}
\newcommand{\nd}{\mathrm{d}}
\newcommand{\nds}{\hat{\mathrm{d}}}

\newcommand{\mG}{\mathcal{G}}

\newcommand{\nG}{\mathrm{G}}
\newcommand{\ep}{\epsilon}

\allowdisplaybreaks

\DeclareMathOperator{\Res}{Res}

\begin{document}

\ensubject{subject}

\ArticleType{Article}
\SpecialTopic{SPECIAL TOPIC: }
\Year{ 2023 }
\Month{ --- }
\Vol{--}
\No{ -- }
\DOI{??}
\ArtNo{-----}
\ReceiveDate{-------}
\AcceptDate{-----, 2023}

\title{Intersection theory rules symbology}{Intersection theory rules symbology}

\author[1]{Jiaqi Chen}{{jiaqichen@csrc.ac.cn}}%
\author[2,1,3]{Bo Feng}{fengbo@csrc.ac.cn}
\author[2]{Li Lin Yang}{yanglilin@zju.edu.cn}

\AuthorMark{J. Chen}

\AuthorCitation{J. Chen, B. Feng, and L. L. Yang}

\address[1]{Beijing Computational Science Research Center, Beijing 100084, China;}
\address[2]{Zhejiang Institute of Modern Physics, School of Physics, Zhejiang University, Hangzhou 310027, China}
\address[3]{Peng Huanwu Center for Fundamental Theory, Hefei, Anhui 230026, China}


\abstract{We propose a novel method to determine the structure of symbols for any family of polylogarithmic Feynman integrals. Using the $\nd\log$-bases and simple formulas for the leading order and next-to-leading contributions to the intersection numbers, we give a streamlined procedure to compute the entries in the coefficient matrices of canonical differential equations, including the symbol letters and the rational coefficients. We also provide a selection rule to decide whether a given matrix element must be zero. The symbol letters are deeply related to the poles of the integrands and also have interesting connections to the geometry of Newton polytopes.
{  Our method can be applied to many cutting-edge multi-loop calculations.}
The simplicity of our results also hints at the possible underlying structure in perturbative quantum field theories.}



\maketitle


\begin{multicols}{2}
\section{Introduction}

Perturbative quantum field theories (pQFTs) play a pivotal role in high-precision phenomenology of high energy physics. In many perturbative calculations, one encounters a class of analytic functions called multiple polylogarithms (MPLs) \cite{Chen:1977oja, Goncharov:1998kja}. They can be mapped to symbols \cite{Goncharov:2010jf, Duhr:2011zq}, which are sequences of $\nd \log W_i$, where the $W_i$'s are algebraic functions of kinematic variables known as symbol letters. For a given scattering process, the complete set of symbol letters is called the ``alphabet''. The knowledge of the alphabet can be used to bootstrap multi-loop integrals and amplitudes \cite{Gaiotto:2011dt, Dixon:2011pw, Dixon:2011nj, Brandhuber:2012vm, Dixon:2013eka, Dixon:2014voa, Dixon:2014iba, Drummond:2014ffa, Dixon:2015iva, Caron-Huot:2016owq, Dixon:2016apl, Dixon:2016nkn, Li:2016ctv, Almelid:2017qju, Chicherin:2017dob, Henn:2018cdp, Drummond:2018caf, Caron-Huot:2019vjl, Caron-Huot:2020bkp, Dixon:2020cnr, Dixon:2020bbt, Guo:2021bym, He:2021eec, Dixon:2022rse, Dixon:2022xqh}. This has stimulated extensive research on the construction of symbol alphabets \cite{Caron-Huot:2011dec, Golden:2013xva, Panzer:2014caa, Dennen:2015bet, Caron-Huot:2016owq, Mago:2020kmp, Abreu:2021vhb, Gong:2022erh, Yang:2022gko, He:2021non, He:2021eec, He:2022tph, He:2023qld}.
In particular, the symbol letters of one-loop integrals have been fully understood \cite{Arkani-Hamed:2017ahv, Abreu:2017enx, Abreu:2017mtm, Chen:2022fyw, Dlapa:2023cvx, Jiang:2023qnl}.
{  However, beyond one-loop, there are no general results available.}
On the other hand, from experiences in multi-loop calculations, the expressions of the symbol letters usually turn out to be much simpler than those in the intermediate steps of the calculations. Hence, in addition to the phenomenological motivations, it is also theoretically interesting to investigate the source of such simplicity and to ask whether it implies the existence of simpler rules for symbology.

The symbols in a polylogarithmic integral family are deeply related to the method of canonical differential equations (CDEs) \cite{Kotikov:1990kg, Kotikov:1991pm, Gehrmann:1999as, Bern:1993kr, Henn:2013pwa}.  This method has become the most streamlined approach to obtain analytic expressions of Feynman integrals. {  One chooses a canonical basis of master integrals with uniform transcendentality (UT) \cite{Henn:2013pwa}, and derive their differential equations with the help of integration-by-parts (IBP) reduction \cite{Chetyrkin:1981qh}. These differential equations are $\epsilon$-factorized (where $d=4-2\epsilon$ in dimensional regularization) and are dubbed ``canonical''.} The entries of the coefficient matrix, if can be written as total derivatives, directly give the symbol letters. The symbols of the solutions to the CDEs can then be iteratively obtained order-by-order in $\ep$. However, converting the coefficient matrix elements to total derivatives can be rather challenging in multivariate situations. Moreover, the procedure of performing the IBP reduction and deriving the CDEs offers little insight into the origin of the symbol letters.

The method of intersection theory \cite{Mastrolia:2018uzb, Frellesvig:2019uqt, Mizera:2019ose, Chestnov:2022xsy} provides an alternative way to reduce the Feynman integrals to master integrals. It is also useful in the construction of UT bases with $\nd\log$-form integrands \cite{Arkani-Hamed:2012zlh,Henn:2013pwa, Chen:2020uyk, Chen:2022lzr, Chicherin:2018old, Bern:2014kca, Henn:2020lye, Dlapa:2021qsl} in the Baikov representation \cite{Baikov:1996iu}. In both the computation of intersection numbers and the construction of UT bases, information of poles in the integrands plays a crucial role. In this Letter, we show that the information of poles also determines the symbol letters to a certain extent. We employ the method of computing intersection numbers from higher-order partial differential equations \cite{Chestnov:2022xsy}, and apply it to the differential equations of the UT bases of \cite{Chen:2020uyk, Chen:2022lzr}. We find that with the universal formulas of the leading order (LO) and next-to-leading order (NLO) contributions to intersection numbers, the symbol letters can be generated by localizing the $\nd\log$-integrands to the multivariate poles. We provide a streamlined procedure to derive all symbol letters in an integral family, that involves the factorization of degenerate poles, followed by simple algebraic operations. {  This can be applied to many cutting-edge multi-loop calculations in pQFTs.}

\section{Symbols from intersection theory}

The Feynman integrals in the Baikov representation are hypergeometric functions of the form 
\begin{equation}
	I[u,\varphi] \equiv \int  u \, \varphi \,, 
\end{equation}
where
\begin{equation}
    u=\prod_i \left[P_i(\bm{z})\right]^{\beta_i} \,,
    \quad
    \varphi \equiv \hat{\varphi}(\bm{z}) \bigwedge_j \nd z_j =\frac{Q(\bm{z})}{\prod_i P_i^{a_i}} \bigwedge_j \nd z_j \,. \label{eq:baikov}
\end{equation}
The sequence of Baikov variables is denoted by $\bm{z}=(z_1,\ldots,z_n)$.
The polynomials $P_i(\bm{z})$ include the Baikov variables themselves, and the Gram determinants $\nG(\bm{q})\equiv\det(q_i \cdot q_j)$ of loop and external momenta.
{  The exponents take the general form $\beta_i = n_i + m_i\epsilon + l_i \delta_i$, where $n_i,m_i,l_i$ are rational numbers, $\epsilon$ is the dimensional regulator, and $\delta_i$ is an optional extra regulator. One usually needs to introduce $\delta_i$ into the computation if $m_i=0$, $n_i$ is integer and $P_i$ appears in the denominator, e.g., when $P_i$ is an inverse propagator.}
The numerator $Q(\bm{z})$ is an arbitrary polynomial of $\bm{z}$.
{  All integrals with the same $u$ form an integral family, within which one can define IBP-equivalence classes of cocycles \cite{Mastrolia:2018uzb, Frellesvig:2019uqt, Mizera:2019ose, Chestnov:2022xsy}:
\begin{equation}
\bra{\varphi_L} \equiv \varphi_L \sim \varphi_L + \sum_i \nabla_i \xi_i \,, \, \quad \nabla_{i} = dz_i \wedge (\partial_{z_i} + \hat{\omega}_i) . \label{eq:braket}
\end{equation}}
where $\omega \equiv \sum_i \hat{\omega}_i \nd z_i$ with $\hat{\omega}_i \equiv \partial_{z_i} \log(u)$.
The dual space consists of equivalence classes $\ket{\varphi}$ of integrals $I[u^{-1},\varphi]$.
The intersection number between $\bra{\varphi_L}$ and $\ket{\varphi_R}$ is given by
\begin{equation}
	\braket{\varphi_L|\varphi_R} = \sum_{\bm{p}} \Res_{\bm{z} = \bm{p}} \left(\psi_L \hat{\varphi}_R\right) , \label{eq:intnum}
\end{equation}
where $\psi_L$ is a function satisfying $\nabla_n \cdots \nabla_1 \psi_L = \varphi_L$. The summation goes over all $n$-variable poles $\bm{p}$ determined by the zeros of the polynomial factors $P_i$ in $u$ \cite{Larsen:2017aqb}. One complication is that some of these poles can be non-factorized, such that the residue can not be computed variable-by-variable in terms of $z_i$. A non-factorized pole can also be degenerate, roughly meaning that more than $n$ factors vanish at the pole.
A simple example is $u=z_1^{\beta_1} z_2^{\beta_2} (z_1+z_2)^{\beta_3}$, for which the pole $\bm{p}=(0,0)$ is non-factorized and degenerate.

To compute the multivariate residues in the presence of non-factorized poles, one can carry out a factorization procedure \cite{Chestnov:2022xsy}. The idea is similar in spirit to the method of sector decomposition \cite{Binoth:2000ps, Binoth:2003ak, Binoth:2004jv, Heinrich:2008si, Bogner:2007cr, Smirnov:2021rhf, Borowka:2017idc}. This involves a change of variables (labelled by $(\alpha)$) from $\bm{z}$ to $\bm{x}^{(\alpha)}$, such that the pole at $\bm{z}=\bm{p}$ corresponds to $\bm{x}^{(\alpha)}=\bm{\rho}^{(\alpha)}$, and in the vicinity of the pole
\begin{equation}
    u(\bm{x}^{(\alpha)})\big|_{\bm{x}^{(\alpha)} \to \bm{\rho}^{(\alpha)}} = \bar{u}_\alpha(\bm{\rho}^{(\alpha)}) \prod_i \left[ x_i^{(\alpha)} - \rho_i^{(\alpha)} \right]^{\gamma_i^{(\alpha)}} ,
    \label{eq:degenerate}
\end{equation}
where $\bar{u}_\alpha(\bm{\rho}^{(\alpha)})$ is non-vanishing. This expression defines the $\bm{u}$-\textbf{powers} $\gamma^{(\alpha)}_i$ for the variable change $(\alpha)$. Note that for each degenerate pole $\bm{p}$, one usually needs to sum over several different factorization to correctly reproduce the multivariate residue. To unify the notation, we also give a label $(\alpha)$ for already factorized poles. Hence the summation in Eq.~\eqref{eq:intnum} is replaced by a summation over $\alpha$ with the residue at $\bm{x}^{(\alpha)}=\bm{\rho}^{(\alpha)}$.
For the simple example $u=z_1^{\beta_1} z_2^{\beta_2} (z_1+z_2)^{\beta_3}$, one possible variable change is $z_1=x_1, z_2 = x_1(x_2-1)$. This leads to $u=x_1^{\beta_1+\beta_2+\beta_3} x_2^{\beta_3} (x_2-1)^{\beta_2}$. We will discuss more about the factorization of degenerate poles later.

With above discussions, we now study the CDE satisfied by a $\nd\log$ basis $\{\bra{\varphi_I}\}$ constructed using the method of \cite{Chen:2020uyk,Chen:2022lzr}.
For later convenience, we choose to keep only the dimensional regulator $\epsilon$ and the  regulators $\delta_i$ for propagators in $\beta_i$ of Eq.~\eqref{eq:baikov}, and absorb all other powers into $a_i$ in $\varphi$. There are two types of building blocks (which will be called the ``rational-type'' and the ``sqrt-type'' in the following):
\begin{align}
	&\nd \log(z-c) = \frac{d z}{z-c}\,, \nn \\
	&\nd \log(\tau[z,c;c_{\pm}]) = \frac{\sqrt{(c-c_+)(c-c_-)}d z}{(z-c)\sqrt{(z-c_+)(z-c_-)}} \,,\nn\\
	&{  \tau[z,c;c_{\pm}]} \equiv \frac{\sqrt{c-c_+} \sqrt{z-c_-}+\sqrt{c-c_-} \sqrt{z-c_+}}{\sqrt{c-c_+}
		\sqrt{z-c_-}-\sqrt{c-c_-} \sqrt{z-c_+}}\,,\label{eq:dlog_basis}
\end{align}
where $c$ and $c_\pm$ are independent of $z$.
The CDE is
\begin{equation}
	\bra{\dot{\varphi}_I} \equiv {  \bra{\nds\varphi_I+ \varphi_I \nds\log u} } = \big( \nds \Omega \big)_{IJ} \bra{\varphi_J} \,,
\end{equation}
where $\nds$ denotes the total derivative with respect to external parameters, such as masses and scalar products (to distinguish, $\nd$ is used for integration variables $\bm{z}$). The matrix $\nds \Omega$ contains all information about the symbol letters, and can be computed by intersection numbers:
\begin{equation}
    \big( \nds \Omega \big)_{IK} = \braket{\dot{\varphi}_I|\varphi_J} \big( \eta^{-1} \big)_{JK} \, ,
    \label{eq:intnumDE}
\end{equation}
where $\eta^{-1}$ is the inverse of the matrix $\eta$ with elements $\eta_{IJ}=\braket{\varphi_I|\varphi_J}$. Apparently, $\big( \nds \Omega \big)_{IK}$ can be nonzero only if there exist at least one $J$ such that the two factors in the above formula are both nonzero.
{  Note that the $\nds \Omega$ matrix is independent of the choice of the ket-basis. Here, we choose the ket-basis with the same representatives as the bra-basis. This choice is convenient for computing intersection numbers, and also helps to reveal the selection rule to be discussed later.}

We now consider the contributions from the factorized pole $\bm{x}^{(\alpha)}=\bm{\rho}^{(\alpha)}$ to the intersection number $\braket{\varphi_L|\varphi_R}$. Around the pole, an $n$-form $\varphi$ can be Laurent-expanded and organized by the powers $\bm{b}=(b_1,\ldots,b_n)$. Such a term can be written as
\begin{equation}
    \varphi^{(\bm{b})} = C^{(\bm{b})} \bigwedge_i  \left[ x_i^{(\alpha)} - \rho_i^{(\alpha)} \right]^{b_{i}} \nd x^{(\alpha)}_i \,. \label{eq:phib}
\end{equation}
In the computation of the intersection number, we know that the contributing terms must have $b_{L,i}+b_{R,i} \leq -2$ for all $i$.
A key point is that a $\nd \log$-form $\varphi_I$ or $\varphi_J$ exhibit only multivariate simple poles, i.e., $b_{i} \geq -1$. The action of $\nds$ may generate terms with one $b_i = -2$. Hence, we only need to consider two kinds of contributions. The LO contribution \cite{ Mizera:2019ose} has all $b_{L,i}+b_{R,i}=-2$, and can be written as
\begin{equation}
	\frac{C_L^{(\bm{b}_L)} C_R^{(\bm{b}_R)}}{ \tilde{\gamma}^{(\alpha)}_1 \cdots \tilde{\gamma}^{(\alpha)}_n} \,, \label{eq:LP}
\end{equation}
where $\tilde{\gamma}^{(\alpha)}_i=\gamma^{(\alpha)}_i - b_{R,i} - 1$.
The NLO contribution has one $b_{L,j}+b_{R,j}=-3$ and the other $b_{L,i}+b_{R,i}=-2$, and can be written as
\begin{equation}
	-\frac{C_L^{(\bm{b}_L)} C_R^{(\bm{b}_R)}}{\tilde{\gamma}^{(\alpha)}_1 \cdots \tilde{\gamma}^{(\alpha)}_n} \, \frac{\partial_{\rho^{(\alpha)}_j} \log \left(\bar{u}_\alpha(\bm{\rho}^{(\alpha)})\right)}{\tilde{\gamma}_j^{(\alpha)}-1 } \,. \label{eq:NLP}
\end{equation}
Here and in the following, we treat each component of $\bm{\rho}^{(\alpha)}$ as an independent external variable. They will be set to the actual expressions in the symbol letters. {  The detailed derivation of the above results are given in \ref{app1}.}

From the above discussion, one finds that $\braket{\dot{\varphi}_I|\varphi_J}$ can receive nonzero contributions from the pole $\bm{x}^{(\alpha)}=\bm{\rho}^{(\alpha)}$ only if $\bm{b}_I$ and $\bm{b}_J$ satisfy either of the following two conditions.  The first condition is that one component $b_{I,k}+b_{J,k}=-1$, while all other $b_{I,i}=b_{J,i}=-1$. We say that $\varphi_I$ and $\varphi_J$ share the $(n-1)$-variable simple pole ($(n-1)$-SP) for $\bm{x}^{(\alpha)}_{\hat{k}}$, where the subscript $\hat{k}$ means that the $k$th variable is removed from the sequence. The derivative $\partial_{\rho^{(\alpha)}_k}$ in $\nds$ generates LO contributions to $\braket{\dot{\varphi}_I|\varphi_J}$.
{  Since $C_I^{(\bm{b}_I)}$ and $C_J^{(\bm{b}_J)}$ may also depend on $\rho^{(\alpha)}_k$, we need to perform an integration to get a total derivative. The contribution can then be be written as}
\begin{equation}
    -{\frac{{\gamma}^{(\alpha)}_k }{\bm{{\gamma}}^{(\alpha)}}} \, \nds \int C_I^{(\bm{b}_I)}C_J^{(\bm{b}_J)} \, \nds \rho^{(\alpha)}_k \,,
    \label{eq:NMCSPdlog}
\end{equation}
where $\bm{\gamma}^{(\alpha)} \equiv \gamma^{(\alpha)}_1 \cdots \gamma^{(\alpha)}_n$. {  Note that here $\gamma_i = \tilde{\gamma}_i$ for $i \neq k$.} In many cases, the product $C_I^{(\bm{b}_I)}C_J^{(\bm{b}_J)}$ is proportional to $(\rho^{(\alpha)}_k-c)^{-1}$, and the integral simply gives rise to the letter $\log(\rho^{(\alpha)}_k-c)$. The more complicated situations will be discussed in the next Section, where we will show that the letters can always be obtained via purely algebraic operations without performing any integration.

The second condition for a nonzero contribution to  $\braket{\dot{\varphi}_I|\varphi_J}$ is $\bm{b}_I=\bm{b}_J=\bm{-1}$, i.e., $\varphi_I$ and $\varphi_J$ share the $n$-variable simple pole ($n$-SP) for $\bm{x}^{(\alpha)}$. The derivative $\nds$ now generates NLO contributions. Using Eq.~\eqref{eq:NLP}, the sum of the NLO contributions are
\begin{equation}
	\frac{C_I^{(\bm{-1})} C_J^{(\bm{-1})}}{\bm{\gamma}^{(\alpha)}} \, \nds \log \Big( \bar{u}_\alpha(\bm{\rho}^{(\alpha)}) \Big) \,.  \label{eq:CSPdlog}
\end{equation}
Note that $\bar{u}_\alpha(\bm{\rho}^{(\alpha)})$ still contains powers $\beta_i$. After taking the $\nds\log$, they become coefficients in front, and the remaining arguments of the logarithms are the symbol letters. We also note that if $u$ has any $\bm{z}$-independent constant factors such as $P_0^{\beta_0}$, it is automatically included in $\bar{u}_\alpha$.

We now turn to the matrix element $\eta_{IJ}=\braket{\varphi_I|\varphi_J}$, which receives LO contributions \eqref{eq:LP} if and only if $\varphi_I$ and $\varphi_J$ share an $n$-SP for at least one factorization $(\alpha)$ (hence, $\eta_{II}$ is always nonzero). To understand when does $(\eta^{-1})_{IJ}\neq 0$, we introduce the concept of $n$-SP chains. If $\varphi_I$ and $\varphi_J$ share an $n$-SP, we say that they are $n$-SP related (denoted as $\varphi_I \sim \varphi_J$). If $\varphi_I \sim \varphi_K$ and $\varphi_I \sim \varphi_J$, the three $n$-forms belong to an $n$-SP chain. This concept straightforwardly generates to more than three $n$-forms. One can see that if $\varphi_I$ and $\varphi_J$ do not belong to an $n$-SP chain, then $(\eta^{-1})_{IJ}= 0$
\footnote{$(\eta^{-1})_{IJ}$ is proportional to the $IJ$-minor of $\eta$. If $\varphi_I$ and $\varphi_J$ do not belong to an $n$-SP chain, all terms in the minor vanish.}.

Combining the condition for nonzero $(\eta^{-1})_{IJ}$ and that for nonzero $\braket{\dot{\varphi}_I|\varphi_J}$, we arrive at the \textbf{selection rule} for nonzero entries in $\nds \Omega$: $(\nds \Omega)_{IJ}$ can be nonzero only if there exists at least one $\varphi_K$ belonging to an $n$-SP chain with $\varphi_J$, and sharing at least one $n$-SP or $(n-1)$-SP with $\varphi_I$. This selection rule, together with the expressions \eqref{eq:NMCSPdlog} and \eqref{eq:CSPdlog} of the symbol letters, serve as the most important results of this paper.

Before closing this section, we show from our results that the differential equation of the $\nd\log$-basis is indeed canonical. Let us assign a transcendental weight of $-1$ to $\beta_i$  (contains $\epsilon$ and $\delta_i$) in Eq.~\eqref{eq:baikov}. Then, all $\gamma^{(\alpha)}_i$ in \eqref{eq:degenerate}
have weight $-1$. Since $\eta_{IJ}$ has the form of Eq.~\eqref{eq:LP} (with $\tilde{\gamma}^{(\alpha)}_i=\gamma^{(\alpha)}_i$), the $\eta^{-1}$ has weight $-n$. Eqs.~\eqref{eq:NMCSPdlog} and \eqref{eq:CSPdlog} have the form of a weight-$(n-1)$ coefficient times a weight-$1$ $\nds \log$. Using Eq.~\eqref{eq:intnumDE}, one can see that $(\nds\Omega)_{IJ}$ is a weight-$(-1)$ coefficient times a $\nds \log$. Hence, we have proved that $\nds\Omega$ is proportional to $\epsilon$ when the regulators $\delta_i$ are taken to zero.

\section{Structure of symbol letters}

We now consider the computations of $\braket{\dot{\varphi}_I|\varphi_J}$ leading to Eqs.~\eqref{eq:NMCSPdlog} and \eqref{eq:CSPdlog} from a different perspective. The expansion of $\varphi_I$ and $\varphi_J$ in the form of Eq.~\eqref{eq:phib} helps to take the $n$-variable residue at once. However, we can always choose to take the $(n-1)$-variable residue of $\bm{x}^{(\alpha)}_{\hat{k}}$ first using Eq.~\eqref{eq:LP}, and leave the dependence on $x^{(\alpha)}_k$ un-expanded. The leftover $1$-form of $x^{(\alpha)}_k$ is a univariate $\nd\log$-form. This operation applies to both $(n-1)$-SP contributions (where $k$ is fixed) and the $n$-SP contributions (where one can freely choose any $k$). Hence, the problem with the single variable $z \equiv x^{(\alpha)}_{k}$ lies in all contributions to the symbol letters. In this Section, we work out this univariate problem generically, and reveal the surprisingly simple structure of symbol letters in the meantime. {  Details of the derivation are given in \ref{app2}.}

To warm up, we first consider the case where $c_1 \equiv \rho^{(\alpha)}_k$ is the pole in a rational-type $\nd\log$. In general, there can be further factors involving $z$ after taking the $(n-1)$-variable residues. Without loss of generality, we take
\begin{align}
	u=P_0^{\beta_0} (z-c_1)^{\beta_1} (z-c_2)^{\beta_2} (z-c_3)^{\beta_3} \,.
\end{align}
and more factors can be easily added. The poles $c_\alpha$ do not necessarily correspond to the poles in the original multivariate problem. Nevertheless, we can use the formulas from the previous Section to solve this univariate problem. The poles and the corresponding u-powers are
\begin{equation}
    c_\alpha \in \{c_1,c_2,c_3,\infty\} \,, ~~~~
	\gamma^{(\alpha)} \in \left\{ \beta_1,\beta_2,\beta_3,-{ \sum_{i=1}^3} \beta_i \right\} ,
\end{equation}
with $\alpha=1,2,3,4$. The space has dimension 2, and the $\nd \log$ basis can be constructed as
\begin{equation}
\varphi_I \in \left\{ \frac{\nd z}{z-c_1} ,\, \frac{\nd z}{z-c_2} \right\} .
\end{equation}
Each $\varphi_I$ involves two poles, $c_I$ and $c_4=\infty$. The relevant intersection numbers can be immediately obtained from Eqs.~\eqref{eq:LP}, \eqref{eq:NMCSPdlog} and \eqref{eq:CSPdlog}: 
\begin{align}
    \braket{\dot{\varphi}_I|\varphi_I} &= \sum_{\alpha \neq I} \frac{\gamma^{(\alpha)}}{\gamma^{(I)}} \, \nds\log(c_{I}-c_{\alpha}) + \eta_{II} \beta_0 \, \nds\log P_0 \,, \nn
    \\
    \braket{\dot{\varphi}_I|\varphi_J} &= -\nds\log(c_{I}-c_{J}) +\eta_{IJ} \beta_0 \, \nds\log P_0 \,, \label{eq:1formDE}
\end{align}
and
\begin{equation}
    \eta =
    \begin{pmatrix}
        \frac{1}{\gamma^{(1)}}+\frac{1}{\gamma^{(4)}} & \frac{1}{\gamma^{(4)}}
        \\
        \frac{1}{\gamma^{(4)}} & \frac{1}{\gamma^{(2)}}+\frac{1}{\gamma^{(4)}}
    \end{pmatrix}
    .
\end{equation}
It is interesting to see that, after taking the $(n-1)$-variable residues, each symbol letter is either the difference between two univariate poles, or the constant factor $P_0$ in $u$.

We now turn to the case where $\rho_k^{(\alpha)}$ appears in a sqrt-type $\nd\log$. We take (here we drop the constant factor $P_0$ for simplicity)
\begin{align}
	u=(z-c_1)^{\beta_1} (z-c_2)^{\beta_2} (z-c_+)^{\beta_3}
	(z-c_-)^{\beta_4} \,.
\end{align}
The poles and their corresponding u-powers are
\begin{align}
    c_{\alpha} &\in \{c_1,c_2,\infty,c_+,c_-\} \,,\nn
    \\
    \gamma^{(\alpha)} &\in \left\{ \beta_1,\beta_2,-\sum_i \beta_i,\beta_3,\beta_4 \right\} .
\end{align}
The $\nd \log$ basis $\{\varphi_I\}$ can be constructed as
\begin{align}
    \nd \log \tau[z,c_1;c_{\pm}], \, \nd \log \tau[z,c_2;c_{\pm}], \, \nd \log \tau[z,\infty;c_{\pm}] \,.
\end{align}
Each $\varphi_I$ has only one pole at $c_I$. However, for intersection numbers, the poles at $c_\pm$ can also contribute. We have
\begin{align}
    \braket{\dot{\varphi}_I|\varphi_I} &= \frac{1}{\gamma^{(I)}} \, \nds\log (\bar{u}_I(c_I)) - \nds\log(c_+-c_-) \nn
    \\
    &+ \nds\log (c_I-c_+) + \nds\log (c_I-c_-) \,, \nn
    \\
	\braket{\dot{\varphi}_I|\varphi_J} &= \braket{\dot{\varphi}_J|\varphi_I} = -\nds \log \tau[c_I,c_J;c_\pm] \,. \label{eq:1formsqrtDE}
\end{align}
Again, it is interesting to note that the symbol letters (including those in $\bar{u}_I$) in Eq.~\eqref{eq:1formsqrtDE} takes the form of the difference between two univariate poles, except the last one. However, for the univariate problem, it is always possible to perform a rationalization to get rid of the square-roots in the context of polylogarithmic Feynman integrals. The last letter in \eqref{eq:1formsqrtDE} then becomes one of those in \eqref{eq:1formDE}. In this sense, we arrive at a surprisingly simple \textbf{structure of symbol letters}: all symbol letters (except the constant factors in $u$) are the difference between two univariate poles after taking the $(n-1)$-variable residues.
Combining Eqs.~\eqref{eq:1formDE} and \eqref{eq:1formsqrtDE} with the $(n-1)$-variable residue already obtained, we have completed the derivation of symbol letters.

\section{Factorization of degenerate poles and Newton polytopes}
\label{sec:fac}

As is evident, the first and the most important step of our method is the factorization of degenerate poles. While this can be done algorithmically following sector decomposition, it is instructive to use an example to get some feeling about the procedure. Let's consider the kite topology defined by
\begin{align}
    z_1&=l_1^2-m^2 \,,\; z_2=(l_2-p)^2-m^2 \,,\; z_3=(l_1-l_2)^2 \,,\nn
    \\
    z_4&=l_2^2 \,,\quad z_5=(l_1-p)^2 \,,\quad p^2=s \,.
\end{align}
We impose cut on $z_1, z_2, z_3$, and hence the $u$ function is given by $u = z_4^{\delta_1} z_5^{\delta_2} [\mG(z_4,z_5)]^{-\epsilon}$, with
\begin{multline}
\mG\equiv 4\nG(l_1,l_2,p)\Big|_{z_1=z_2=z_3=0} = - 2m^6 + m^4 (s + z_4 + z_5)
\\
+ m^2 (2 z_4 z_5 - s z_4 - s z_5) + z_4 z_5 (s-z_4-z_5) \,.
\label{eq:Gz4z5}
\end{multline}
From $u$, we can determine the set of poles for $(z_4,z_5)$:
\begin{equation}
    \bm{p} \in \left\{ (0,0), (m^2,m^2), (\infty,0), (0,\infty), (\infty,\infty) \right\} .
\end{equation}
Here we focus on the three-fold degenerate pole $(\infty,0)$.
{  The complete results for this family are given in \ref{app3}.}
For convenience, we first introduce the variable change $z_4 = 1/t_4$, and rewrite $u=t_4^{2\epsilon-\delta_1} z_5^{\delta_2} \, \mG_{\infty 0}^{-\epsilon}$. Here
\begin{equation}
    \mG_{\infty 0} \equiv t_4^2 \, \mG(1/t_4,z_5) \equiv t_4 [r_+(t_4) - z_5] [z_5 - r_-(t_4))] \, ,
    \label{eq:Ginf0}
\end{equation}
where the last equal sign defines the two roots $r_\pm(t_4)$ of $\mG_{\infty 0}$ with respect to $z_5$. Noting that
\begin{equation}
    z_5 - r_-(t_4) = z_5 - m^2(m^2-s)t_4 + \mathcal{O}(t_4^2) \,,
    \label{eq:exproot}
\end{equation}
we find 3 factors in $u$ vanishing when $(t_4,z_5)=(0,0)$: $t_4$, $z_5$ and $z_5-r_-(t_4)$. There are 3 different factorization, corresponding to 3 ways to organize the 3 factors into 2 groups:
\begin{align}
\bm{x}^{(4)} &: (\{t_4\} , \{z_5,z_5-r_{-}(t_4)\}) \,,\nn\\
\bm{x}^{(5)} &: (\{t_4,z_5-r_{-}(t_4)\},\{z_5\}) \,,\nn\\
\bm{x}^{(6)} &: (\{z_5-r_{-}(t_4)\} , \{t_4,z_5\}) \,.
\end{align}
As an example, for $\bm{x}^{(5)}$ we have the variable change
\begin{equation}
    t_4 = x^{(5)}_1 \,, \quad z_5 = x^{(5)}_1 x^{(5)}_2 \, ,
\end{equation}
which leads to (see Eq.~\eqref{eq:degenerate})
\begin{equation}
    \bar{u}_5(\bm{\rho}^{(5)}) = [m^2(m^2-s)]^{-\epsilon} \,, \; \gamma^{(5)}_i \in \{ \epsilon-\delta_1 + \delta_2, \delta_2 \} \,,
\end{equation} 
where $\bm{\rho}^{(5)}=(0,0)$.

The integral family has four master integrals and exhibits a symmetry under $z_4\leftrightarrow z_5$ and $\delta_1\leftrightarrow \delta_2$. The $\nd \log$ basis can be constructed as
\begin{align}
\varphi_{1}&=\frac{\nd z_4 \nd z_5}{z_4 z_5} \,,\quad \varphi_{2}=\frac{\sqrt{s(s-4m^2)}}{\mG}\nd z_4 \nd z_5 \,,\nn\\
\varphi_{3}&=\frac{ z_4-m^2 }{\mG}\nd z_4 \nd z_5 \,, \quad \varphi_{4}=\frac{ z_5-m^2}{\mG} \nd z_4 \nd z_5 \,.
\end{align}
With the variable change to $\bm{x}^{(5)}$ and the expansion around $\bm{\rho}^{(5)}$, the leading terms of $\varphi_1$ and $\varphi_3$ are
\begin{align}
    &\varphi_1^{(-1,-1)} = \frac{\nd x_1^{(5)} \nd x_2^{(5)}}{x_1^{(5)} x_2^{(5)}} \, ,\nn\\
    &\varphi_3^{(-1,0)} = \frac{\nd x_1^{(5)} \nd x_2^{(5)}}{x_1^{(5)} \left[\rho^{(5)}_2 +  m^2(m^2-s) \right]}, ~~~~ \rho^{(5)}_2=0.
\end{align}
Apparently, they share the $(n-1)$-SP for $\bm{x}^{(5)}_{\hat{2}}$. We can then immediately obtain the letter in $\braket{\dot{\varphi}_1|\varphi_3}$ from Eq.~\eqref{eq:NMCSPdlog}, or from Eq.~\eqref{eq:1formDE} as the difference between two univariate poles:
\begin{equation}
    m^2(m^2-s) \, .
    \label{eq:NMCSPeg}
\end{equation}

\begin{figure}[H]
\centering
    \includegraphics[width=2.5in]{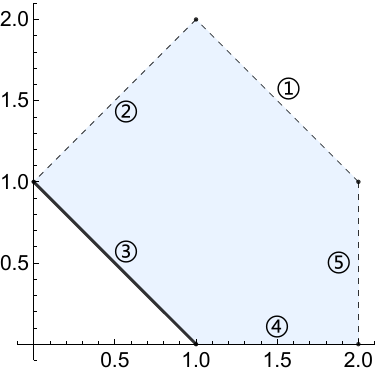}
    \caption{The Newton polytope of $\mG_{\infty 0}$. Horizontal and vertical axis are the power of $t_4$ and $z_5$.}
    \label{fig:NP}
\end{figure}

We now make an interesting observation: the letter in Eq.~\eqref{eq:NMCSPeg} is just the ratio between the coefficient of $t_4$ and that of $z_5$ in Eq.~\eqref{eq:exproot} (as well as in $\mG_{\infty 0}$). {  These two terms are the leading ones in the limit $t_4 \to 0$ and $z_5 \to 0$. Newton polytopes provide a geometric view to study limits of multivariate polynomials. A Newton polytope is the convex hull of the exponent-vectors of a polynomial. This geometric view has been used to study singularities of Feynman integrands. See, e.g., \cite{Arkani-Hamed:2022cqe} for ultraviolet and infrared divergences, \cite{Pak:2010pt} for the method of regions \cite{Beneke:1997zp}, and \cite{Kaneko:2009qx} for sector decomposition \cite{Binoth:2000ps, Binoth:2003ak, Binoth:2004jv, Heinrich:2008si, Bogner:2007cr, Smirnov:2021rhf, Borowka:2017idc}.}
These motivate us to understand the symbol letters from Newton polytopes. The Newton polytope of $\mG_{\infty 0}$ is shown in Fig.~\ref{fig:NP}. It has five facets. Since the components of the outer normal vector of facet \ding{194} are all negative, this facet is degenerate and the corresponding polynomial is exactly Eq.~\eqref{eq:exproot}. The letter \eqref{eq:NMCSPeg} is essentially the ratio of the two coefficients at the vertices of the degenerate facet. The other $(n-1)$-SP contributions of the form $c_I-c_J$ to $\braket{\dot{\varphi}_I|\varphi_J}$ follow a similar pattern.

Similar observations can also be made for the $n$-SP contributions. There are two possibilities here. The first case is when there is a degenerate facet, and then the coefficient at one of its vertices gives the letter. For example, the contribution from $\bm{\rho}^{(4)}$ to $\braket{\dot{\varphi}_1|\varphi_1}$ is given by the vertex $(0,1)$, and the letter is $\nds\log (-1)=0$; while the contribution from $\bm{\rho}^{(5)}$ is given by the vertex $(1,0)$, and the letter is the same as Eq.~\eqref{eq:NMCSPeg}. The second possibility is when there is no degenerate facet. In this case, the origin $(0,0)$ must be a vertex of the polytope, and 
its coefficient gives a letter. For example, the contributions from $\bm{p}=(0,0)$ to $\braket{\dot{\varphi}_1|\varphi_1}$ is related to vertex $(0,0)$ of the polytope corresponding to $\mG(z_4,z_5)$. Hence, the letter is given by the constant term of Eq.~\eqref{eq:Gz4z5}:
\begin{equation}
\mG(0,0) = m^4(s-2 m^2) \,.
\label{eq:G00}
\end{equation}

We have checked that the other contributions do not give rise to new letters, and Eqs.~\eqref{eq:NMCSPeg} and \eqref{eq:G00} are already the full set of letters in this simple example. 

{  The example discussed above is simple with only two integration variables and only involving rational letters depending on two kinematic variables. We emphasize that our method can be applied to problems with more variables and with irrational letters as well. In particular, we have tested our method in multivariate one-loop examples with irrational letters, and find agreement with existing results. Applications in more complicated multi-loop examples are in progress and will be presented in a forthcoming article.}

\section{Summary and outlooks}

In this Letter, we propose a novel method to determine the structure of symbols for any family of polylogarithmic Feynman integrals using intersection theory. The procedure is purely algebraic, involving factorization of degenerate poles and computation of residues at simple poles. The computation of intersection numbers also gives the rational coefficients in the CDEs, and hence completely determines the latter. In particular, we have found a selection rule for nonzero entries in the CDEs.

Our results also reveal some interesting structures underlying the symbol letters. We find that all symbol letters are either the constant factors in the $u$-function, or the differences between univariate poles after taking the residues for the other variables.  We also take a first glance at the possible relationship between the symbol letters and the Newton polytopes
associated with the polynomial factors in the $u$-function. We hope that these algebraic and geometric structures can be used to further simplify the calculation of symbol letters, and provide insights about the mathematical structure of QFT.

In recent years, there have been enormous efforts to extend the concept of pure functions to Feynman integrals beyond the polylogarithmic cases (see, e.g., \cite{Broedel:2018qkq, Frellesvig:2023iwr}). It is interesting to see whether our method can be generalized to those cases as well. Moreover, since differential equations can be regarded as iterative reduction relations \cite{Chen:2022jux}, our result also serves as a development towards simplifying the reduction procedure, and shows the connection between the analytic and algebraic structures of Feynman integrals.

\Acknowledgements{We thank Xuhang Jiang for valuable discussions. This work is supported by Chinese NSF funding under Grant No.11935013, No.11947301, No.12047502 (Peng Huanwu Center), No.12247120, No.12247103, NSAF grant No.U2230402, and China Postdoctoral Science Foundation No.2022M720386, and is supported in part by the National Natural Science Foundation of China under Grant No.11975030 and 12147103, and the Fundamental Research Funds for the Central Universities.}

\InterestConflict{The authors declare that they have no conflict of interest.}


\begin{appendix}





\section{Intersection numbers from factorized poles} \label{app1}

In this Section, we review the calculation of intersection numbers that leads to the leading-order (LO) and next-to-leading order (NLO) contributions \eqref{eq:LP} and \eqref{eq:NLP}. For the moment we will suppress the superscript $(\alpha)$ labeling the factorization transformations, and assume $\bm{x} = \bm{\rho}$ is already a factorized pole. Around this pole, the $u$ function can be written as
\begin{equation}
u(\bm{x}) = \bar{u}(\bm{x}) \prod_i \left[ x_i - \rho_i \right]^{\gamma_i} \,,
\end{equation}
where $\bar{u}(\bm{x})$ can be Taylor-expanded as
\begin{align}
&\bar{u}(\bm{x}) = \bar{u}(\bm{\rho}) + \sum_i (x_i - \rho_i) \left[ \frac{\partial}{\partial x_i} \bar{u}(\bm{x}) \right]_{\bm{x} = \bm{\rho}} \nn\\
&+ \frac{1}{2} \sum_{i,j} (x_i - \rho_i) (x_j - \rho_j) \left[ \frac{\partial^2}{\partial x_i \partial x_j} \bar{u}(\bm{x}) \right]_{\bm{x} = \bm{\rho}} + \cdots \, .
\end{align}
The $n$-form $\varphi_L$ can be similarly decomposed as
\begin{equation}
\varphi_L = \sum_{\bm{b}_L} \varphi_L^{(\bm{b}_L)} \equiv \sum_{\bm{b}_L} C_L^{(\bm{b}_L)} \bigwedge_i \left( x_i - \rho_i \right)^{b_{L,i}} dx_i \, ,
\end{equation}
where $\bm{b}_L = (b_{L,1},\ldots,b_{L,n})$ denotes a vector of powers.
The covariant derivative $\nabla_i$ with respect to $x_i$ is defined as
\begin{equation}
    \nabla_i = dx_i \wedge (\partial_{x_i} + \omega_i) \, ,
\end{equation}
where
\begin{equation}
    \omega_i \equiv \partial_{x_i} \log(u) = \frac{\gamma_i}{x_i - \rho_i} + \partial_{x_i} \log(\bar{u}) \, .
\end{equation}

We now need to look for a function $\psi_L$ which satisfies $\nabla_n \cdots \nabla_1 \psi_L = \varphi_L$ around the pole.
The above equation is linear in $\psi_L$ and $\varphi_L$. Hence we can decompose the solution as
\begin{equation}
    \psi_L = \sum_{\bm{b}_L} \psi_L^{(\bm{b}_L)} \, , \quad \nabla_n \cdots \nabla_1 \psi_L^{(\bm{b}_L)} = \varphi_L^{(\bm{b}_L)} \, .
    \label{eq:psi_equation}
\end{equation}
We can write the Ansatz for $\psi_L^{(\bm{b}_L)}$ as 
\begin{align}
&\psi_L^{(\bm{b}_L)} = C_L^{(\bm{b}_L)} \left[ A^{(0)} + \sum_j A^{(1)}_j (x_j - \rho_j) \right. \nn \\
&\left. ~~ + \frac{1}{2} \sum_{j,k} A^{(2)}_{j,k} (x_j-\rho_j) (x_k-\rho_k) + \cdots \right] \prod_i (x_i   -\rho_i)^{b_{L,i} + 1} .
\end{align}
Plugging the above into Eq.~\eqref{eq:psi_equation}, the covariant derivatives give rise to
\begin{align}
    &\left( \prod_i \nabla_i \right) \psi_L^{(\bm{b}_L)} = C_L^{(\bm{b}_L)} \left[ \prod_i (x_i-\rho_i)^{b_{L,i}} \right] \nonumber
    \\
    & \times \left[ A^{(0)} \prod_i (\gamma_i + b_{L,i} + 1) + \sum_j (x_j-\rho_j) \left[ A_j^{(1)} (\gamma_j + b_{L,j} + 2) \right. \right. \nn \\
    & \left. \left. ~~~ + A^{(0)} \partial_{\rho_j} \bar{u}(\bm{\rho}) \right] \prod_{i \neq j} (\gamma_i + b_{L,i} + 1) + \cdots \right] .
\end{align}
Hence, we find that the coefficients are given by
\begin{align}
    A^{(0)} &= \frac{1}{\prod_i(\gamma_i+b_{L,i}+1)} \,, ~~~~ A^{(1)}_j = -\frac{A^{(0)} \partial_{\rho_j} \log(\bar{u}(\bm{\rho}))}{\gamma_j+b_{L,j}+2}  \,.
\end{align}

It is now straightforward to compute the intersection numbers. Supposing that $\varphi_R$ is given by
\begin{equation}
\varphi_R = \sum_{\bm{b}_R} \varphi_R^{(\bm{b}_R)} \equiv \sum_{\bm{b}_R} C_R^{(\bm{b}_R)} \bigwedge_i \left( x_i - \rho_i \right)^{b_{R,i}} dx_i \, ,
\end{equation}
The contribution from the factorized pole $\bm{x} = \bm{\rho}$ to the intersection number between $\varphi_L^{(\bm{b}_L)}$ and $\varphi_R^{(\bm{b}_R)}$ is given by
\begin{align}
    &\Res_{\bm{x} = \bm{\rho}} \left( \psi_L^{(\bm{b}_L)} \varphi_R^{(\bm{b}_R)} \right) = \Res_{\bm{x} = \bm{\rho}} C_L^{(\bm{b}_L)} C_R^{(\bm{b}_R)} \prod_i (x_i-\rho_i)^{b_{L,i} + b_{R,i} + 1} \nn\\
    &\quad \quad \quad\quad \quad\quad \quad \times\left[ A^{(0)} + \sum_j A^{(1)}_j (x_j - \rho_j) + \cdots \right]  \, .
\end{align}
When $\bm{b}_L + \bm{b}_R = \bm{-2}$, the $A^{(0)}$ term gives rise to the so-called LO contribution (Eq.~\eqref{eq:LP})
\begin{equation}
    \Res_{\bm{x} = \bm{\rho}} \left( \psi_L^{(\bm{b}_L)} \varphi_R^{(\bm{b}_R)} \right) =  \frac{C_L^{(\bm{b}_L)} C_R^{(\bm{b}_R)}}{\prod_i \tilde{\gamma}_i} \,,
    \label{eq:app:first_order}
\end{equation}
where $\tilde{\gamma}_i = \gamma_i - b_{R,i} - 1$. When all $b_{L,i}+b_{R,i} = -2$ except one $b_{L,j}+b_{R,j}=-3$, the $A^{(1)}_j$ term gives rise to the so-called NLO contribution (Eq.~\eqref{eq:NLP})
\begin{equation}
    \Res_{\bm{x} = \bm{\rho}} \left( \psi_L^{(\bm{b}_L)} \varphi_R^{(\bm{b}_R)} \right) =  - \frac{C_L^{(\bm{b}_L)} C_R^{(\bm{b}_R)} \, \partial_{\rho_j} \log(\bar{u}(\bm{\rho})) }{(\gamma_j+b_{L,j}+1) \prod_i \tilde{\gamma}_i}  \,.
    \label{eq:app:second_order}
\end{equation}

At this point, it is worth noting that the contributions in Eqs.~\eqref{eq:app:first_order} and \eqref{eq:app:second_order} are invariant under a simultaneous rescaling of $u$, $\varphi_L^{(\bm{b}_L)}$ and $\varphi_R^{(\bm{b}_R)}$. In terms of the powers $\gamma_i$, $b_{L,i}$ and $b_{R,i}$, this rescaling amounts to the shifts
\begin{equation}
    \gamma_i \to \gamma_i + \xi_i \, , \quad b_{L,i} \to b_{L,i} - \xi_i \, , \quad b_{R,i} \to b_{R,i} + \xi_i \,.
\end{equation}
The shifts do not change the values of $b_{L,i}+b_{R,i}$, $\gamma_i + b_{L,i}$ and $\gamma_i - b_{R,i}$, and hence the expressions for the LO and NLO contributions are manifestly invariant. In the case $\bm{b}_L + \bm{b}_R = \bm{-2}$, we can employ this freedom to make $\bm{b}_L \to \bm{-1}$ and $\bm{b}_R \to \bm{-1}$, i.e., both $\varphi_L^{(\bm{b}_L)}$ and $\varphi_R^{(\bm{b}_R)}$ have only simple poles. The intersection numbers in this situation are well-understood in the literature \cite{Mizera:2019ose}, and agree with Eq.~\eqref{eq:app:first_order}.

As a special case of the above general formulas, we consider the intersection numbers $\braket{\dot{\varphi}_I | \varphi_J}$, where both $\varphi_I$ and $\varphi_J$ are $\nd\log$-forms. We again expand $\varphi_I$ as
\begin{equation}
\varphi_I = \sum_{\bm{b}_I} \varphi_I^{(\bm{b}_I)} \equiv \sum_{\bm{b}_I} C_I^{(\bm{b}_I)} \bigwedge_i \left( x_i - \rho_i \right)^{b_{I,i}} dx_i \, ,
\end{equation}
and similarly for $\varphi_J$. For each $\bm{b}_I$ and $\rho_j$, there is a term in $\dot{\varphi}_I$ given by
\begin{equation}
-(\gamma_j + b_{I,j}) \, \nds\rho_j \; C_I^{(\bm{b}_I)} \bigwedge_i \left( x_i - \rho_i \right)^{b_{I,i}-\delta_{ij}} dx_i \,.
\label{eq:dot_phiI}
\end{equation}
Hence, setting $b_{L,i}=b_{I,i}-\delta_{ij}$ and $b_{R,i}=b_{J,i}$, we can readily use Eqs.~\eqref{eq:app:first_order} and \eqref{eq:app:second_order} to compute the residues. If $\bm{b}_L = \bm{b}_R = \bm{-1}$ (which means $b_{I,j}=0$, i.e., $(n-1)$-SP), the term gives rise to a LO contribution
\begin{equation}
    -\frac{\gamma_j}{\prod_i \gamma_i} C_I^{(\bm{b}_I)}C_J^{(\bm{b}_J)} \, \nds \rho_j \,.
\end{equation}
On the other hand, if $\bm{b}_I = \bm{b}_J = \bm{-1}$ (i.e., $n$-SP), the term leads to a NLO contribution
\begin{equation}
	\frac{C_I^{(\bm{-1})} C_J^{(\bm{-1})}}{\prod_i \gamma_i} \, \partial_{\rho_j} \log(\bar{u}(\bm{\rho})) \, \nds\rho_j \,.  
\end{equation}

\section{Reduction to univariate problems} \label{app2}

In the previous section, we've seen that in the computation of $\braket{\dot{\varphi}_I | \varphi_J}$ for $\nd\log$-forms $\varphi_I$ and $\varphi_J$, the contributing terms $\varphi_I^{(\bm{b}_I)}$ and $\varphi_J^{(\bm{b}_J)}$ share at least $(n-1)$-variable simple poles. Without loss of generality, we denote these $(n-1)$ variables as $\bm{x}_{\hat{1}} = (x_2,\ldots,x_{n})$, and denote the remaining variable as $z \equiv x_1$. In the computation of intersection numbers, one may take the $(n-1)$-variable residues at $\bm{x}_{\hat{1}} = \bm{\rho}_{\hat{1}}$ first, and deal with the single variable $z$ in the last step.

To see how that works, we assume that both $\varphi_L$ and $\varphi_R$ have simple poles at $\bm{x}_{\hat{1}} = \bm{\rho}_{\hat{1}}$. They can then be written as
\begin{align}
    &\varphi_L = f_L(z,\bm{x}_{\hat{1}}) \, dz \bigwedge_{i=2}^{n} \left( x_i - \rho_i \right)^{-1} dx_i \, , \nn\\
    &\varphi_R = f_R(z,\bm{x}_{\hat{1}}) \, dz \bigwedge_{i=2}^{n} \left( x_i - \rho_i \right)^{-1} dx_i \, ,
\end{align}
where $f_L$ and $f_R$ are regular at $\bm{x}_{\hat{1}} = \bm{\rho}_{\hat{1}}$. The $u(\bm{x})$ function can also be written as
\begin{equation}
u(\bm{x}) = \bar{u}(z,\bm{x}_{\hat{1}}) \prod_{i=2}^{n} \left[ x_i - \rho_i \right]^{\gamma_i} \,.
\end{equation}
To compute $\braket{\varphi_L | \varphi_R}$, we need to find a $\psi_L$ satisfying $\nabla_n \cdots \nabla_1 \psi_L = \varphi_L$ in the vicinity of the pole. Due to the simple pole structure, it is straightforward to perform the inversion of $\nabla_i$ for $i=2,\ldots,n$. This leads to 
\begin{equation}
    \nabla_1 \psi_L = \frac{f_L(z,\bm{\rho}_{\hat{1}}) \, dz}{\gamma_2 \cdots \gamma_n} + \mathcal{O}((\bm{x}_{\hat{1}}-\bm{\rho}_{\hat{1}})^0) \, ,
\end{equation}
where the higher-power terms do not contribute since $\varphi_R$ has simple poles. Hence, the computation of the $n$-variable intersection number is equivalent to a univariate problem with
\begin{equation}
    u(z) \equiv \bar{u}(z,\bm{\rho}_{\hat{1}}) \, , \quad \varphi_L \equiv \frac{f_L(z,\bm{\rho}_{\hat{1}}) \, dz}{\gamma_2 \cdots \gamma_n} \,, \quad \varphi_R \equiv f_R(z,\bm{\rho}_{\hat{1}}) \, dz \,.
\end{equation}
Now, we may collect all contributions to $\braket{\dot{\varphi}_I | \varphi_J}$ from the $(n-1)$-variable simple pole at $\bm{x}_{\hat{1}} = \bm{\rho}_{\hat{1}}$ and an additional pole (not necessarily simple) for the variable $z=x_1$. This allows us to study the symbol letters using only univariate $\nd\log$-constructions and intersection numbers.

We first look at the case of rational-type $\nd\log$-forms. The $u$-function can be factorized into
\begin{align}
	u(z) = P_0^{\beta_0} \prod_{\alpha=1}^{\nu+1} (z-c_\alpha)^{\beta_\alpha} \,.
\end{align}
There are $\nu+2$ different poles for $z$
\begin{align}
    \rho^{(\alpha)} &\in \left\{ c_1, \ldots, c_{\nu+1}, \infty \right\} , ~~~~
    \gamma^{(\alpha)} \in \left\{ \beta_1, \ldots, \beta_{\nu+1}, -\sum_{\alpha=1}^{\nu+1} \beta_\alpha \right\} .
\end{align}
For this $u$-function, there are $\nu$ independent integrands. They can be chosen as $\varphi_I = dz/(z-c_I)$ for $I=1,\ldots,\nu$. We need to consider two kinds of intersection numbers: $\braket{\dot{\varphi}_I|\varphi_I}$ and $\braket{\dot{\varphi}_I|\varphi_J}$ with $I \neq J$. For the first kind, we take $\braket{\dot{\varphi}_1|\varphi_1}$ as an example. For that we need to consider $\partial_{\rho^{(1)}} \varphi_1$, $\partial_{\rho^{(\alpha)}} \varphi_1$ for $\alpha \neq 1$, and the symbol letters contained in $P_0$. Here with an abuse of the notation, $\partial_\rho \varphi$ actually denotes $\partial_\rho (u \, \varphi) / u$. Using the formulas for LO and NLO contributions to intersection numbers, we have
\begin{align}
    &\Braket{\partial_{c_1}\varphi_1|\varphi_1} = \Braket{ \frac{(1-\beta_1) \, \nd z}{(z-c_1)^2} | \frac{\nd z}{z-c_1} } = \frac{1}{\beta_1} \, \partial_{c_1} \log \left(\bar{u}_1(c_1)\right) , \nn
    \\
    &\Braket{\partial_{c_{\alpha}} \varphi_1|\varphi_1} = \Braket{\frac{\beta_\alpha \, \nd z}{(z-c_1)(c_\alpha-z)} | \frac{\nd z}{z-c_1}} \nn\\
    &\quad \quad \quad = \frac{\beta_\alpha}{\beta_1} \, \partial_{c_\alpha} \log (c_1-c_\alpha) = \frac{1}{\beta_1} \, \partial_{c_\alpha} \log \left(\bar{u}_1(c_1)\right) ,
\end{align}
where
\begin{equation}
    \bar{u}_1(z) = P_0^{\beta_0} \prod_{\alpha=2}^{\nu+1} (z-c_\alpha)^{\beta_\alpha} \,.
\end{equation}
From the above results, one may easily reconstruct $\braket{\dot{\varphi}_1|\varphi_1}$ in the form of $\nds \log$s, which coincides with Eq.~\eqref{eq:CSPdlog} and the first line of Eq.~\eqref{eq:1formDE}.
For $\braket{\dot{\varphi}_I|\varphi_J}$, we only need to consider the contributions from $\partial_{c_I}\varphi_I$ and $\partial_{c_J}\varphi_I$, as well as from $P_0$. Using the formula for LO intersection numbers, we have
\begin{align}
\Braket{\partial_{c_I} \varphi_I|\varphi_J} &= (1-\beta_I)  \Braket{\frac{\nd z}{(z-c_I)^2} | \frac{\nd z}{z-c_J}} \nn\\
&= - \partial_{c_I} \log (c_I-c_J) \,, \nn
\\
\Braket{\partial_{c_J} \varphi_I|\varphi_J} &= -\beta_J  \Braket{\frac{\nd z}{(z-c_I)(z-c_J)} | \frac{\nd z}{z-c_J}} \nn\\
&= - \partial_{c_J} \log (c_I-c_J) \,.
\end{align}
These agree with the results in Eq.~\eqref{eq:NMCSPdlog} and the second line of Eq.~\eqref{eq:1formDE}

We now move to sqrt-type $\nd\log$-forms. The $u$-function is given by
\begin{align}
	u(z) = P_0^{\beta_0} (z-c_+)^{\beta_+} (z-c_-)^{\beta_-} \prod_{\alpha=1}^{\nu-1} (z-c_\alpha)^{\beta_\alpha} \,.
\end{align}
There are again $\nu+2$ different poles for $z$
\begin{align}
    \rho^{(\alpha)} &\in \left\{ c_1, \ldots, c_{\nu-1}, \infty, c_+, c_- \right\} , \nn
    \\
    \gamma^{(\alpha)} &\in \left\{ \beta_1, \ldots, \beta_{\nu-1}, -\sum_{\alpha=1}^{\nu-1} \beta_\alpha - \beta_+ - \beta_-, \beta_+, \beta_- \right\} .
\end{align}
The two poles $c_\pm$ are singled out to remind us that there is always a factor of $\sqrt{(z-c_+)(z-c_-)}$ in the integrands according to the second equation in Eq.~\eqref{eq:dlog_basis}, which we reproduce here:
\begin{align}
    \nd \log(\tau[z,c;c_{\pm}]) &\equiv \nd \log \frac{\sqrt{c-c_+} \sqrt{z-c_-}+\sqrt{c-c_-} \sqrt{z-c_+}}{\sqrt{c-c_+} \sqrt{z-c_-}-\sqrt{c-c_-} \sqrt{z-c_+}}
    \nn\\
    &= \frac{\sqrt{(c-c_+)(c-c_-)} \, \nd z}{(z-c)\sqrt{(z-c_+)(z-c_-)}} \,.
\end{align}
At this point, we note that the square root of a linear function is related to that of a quadratic function via a variable change. For example, setting $z = 1/t + c_+$, we have
\begin{equation}
    \frac{dz}{\sqrt{(z-c_+)(z-c_-)}} = \frac{dt}{t \, \sqrt{1+t\,(c_+-c_-)}} \,.
\end{equation}
Hence, we don't have to consider the linear function case separately.

For each $I=1,\ldots,\nu-1$, there is an independent integrand $\varphi_I = \nd\log(\tau_I) \equiv \nd \log \tau[z,c_I;c_{\pm}]$. The $\nu$th independent integrand is associated with the pole $\rho^{(\nu)} = \infty$, and is given by
\begin{align}
    \varphi_\nu &= \nd\log(\tau_\nu) \equiv \nd\log \tau[z,\infty;c_\pm] \nn\\
    &= \nd\log \frac{\sqrt{z-c_-}+\sqrt{z-c_+}}{\sqrt{z-c_-}-\sqrt{z-c_+}} 
    = \frac{\nd z}{\sqrt{(z-c_+)(z-c_-)}} \,.
    \label{eq:tau_nu}
 \end{align}
The intersection numbers $\braket{\dot{\varphi}_I | \varphi_J}$ can now be computed as usual. Taking $\braket{\dot{\varphi}_1 | \varphi_1}$ as an example. We need to consider the derivatives with respect to $c_1$, $c_\pm$ and $c_\alpha$ for $\alpha=2,\ldots,\nu-1$. We have
\begin{align}
    \partial_{c_1} \varphi_1 &= \frac{1-\beta_1}{(z-c_1)^2} + \frac{\beta_1}{2} \left[ \frac{1}{c_1-c_+} + \frac{1}{c_1-c_-} \right] \frac{1}{z-c_1} \nn\\
    &+ \mathcal{O}\left((z-c_1)^0\right) , \nn
    \\
    \varphi_1 &= \frac{1}{z-c_1} - \frac{1}{2} \left[ \frac{1}{c_1-c_+} + \frac{1}{c_1-c_-} \right] + \mathcal{O}\left((z-c_1)^1\right) , \nn
    \\
    \partial_{c_\pm} \varphi_1 &= \frac{1/2-\beta_\pm}{z-c_\pm} \, \varphi_1 \,, \nn
    \\
    \partial_{c_\alpha} \varphi_1 &= -\frac{\beta_\alpha}{z-c_\alpha} \, \varphi_1 \,.
\end{align}
There are two terms in $\partial_{c_1} \varphi_1$, leading to both LO and NLO contributions from the pole $c_1$ to the intersection number:
\begin{align}
    \Braket{\partial_{c_1} \varphi_1 | \varphi_1} &= \frac{1}{\beta_1} \partial_{c_1} \log\bar{u}_1(c_1) \nn\\
    & + \partial_{c_1} \log(c_1-c_+) + \partial_{c_1} \log(c_1-c_-) \,.
\end{align}
The intersection number $\braket{\partial_{c_\pm} \varphi_1 | \varphi_1}$ receive LO contributions from the pole $c_1$ as well as $c_\pm$, which are given by
\begin{align}
    \Braket{\partial_{c_\pm} \varphi_1 | \varphi_1} &=  \frac{\beta_\pm - 1/2}{\beta_1} \, \partial_{c_\pm} \log(c_1-c_\pm) \nn\\
    &- \partial_{c_\pm} \log(c_\pm-c_\mp) + \partial_{c_\pm}\log(c_1-c_\pm) \,.
\end{align}
Finally, the intersection number $\braket{\partial_{c_\alpha} \varphi_1 | \varphi_1}$ for $\alpha=2,\ldots,\nu-1$ receive LO contributions only from the $c_1$ pole:
\begin{equation}
\braket{\partial_{c_\alpha} \varphi_1 | \varphi_1} = \frac{\beta_\alpha}{\beta_1} \partial_{c_\alpha} \log(c_1-c_\alpha) = \frac{1}{\beta_1} \partial_{c_\alpha} \bar{u}_1(c_1) \,.
\end{equation}
Combining the above results, we can reproduce the first equation in Eq.~\eqref{eq:1formsqrtDE}. Similarly, $\partial_{c_I}$, $\partial_{c_j}$ and $\partial_{c_\pm}$ give the same contribution as shown in the second equation in Eq.~\eqref{eq:1formsqrtDE}.

Alternatively, one may perform a variable change to rationalize the square root, and compute the intersection numbers in the same way as the rational case. The relevant variable change is simply
\begin{align}
z = \frac{c_+ (\tau_\nu+1)^2 - c_- (\tau_\nu-1)^2}{4 \tau_\nu} \,.
\end{align}
where the variable $\tau_\nu$ is defined in Eq.~\eqref{eq:tau_nu}.
The poles for the new variable $\tau_\nu$ can be written in terms of  a set of new constants
\begin{equation}
    t_I \equiv \tau[\infty,c_I;c_\pm] = \frac{\sqrt{c_I-c_+} + \sqrt{c_I-c_-}}{\sqrt{c_I-c_+} - \sqrt{c_I-c_-}} \,.
\end{equation}
The $\nd\log$ integrands can then be rewritten as $d\log(\tau_\nu)$ and
\begin{align}
\nd\log(\tau_I) = \nd \log (\tau_\nu-t_I) - \nd \log \left(\tau_\nu-\frac{1}{t_I} \right) .
\end{align}
As promised, all integrands are of the rational-type, and the symbol letters can be read off using the existing results.

\section{Details of the kite topology} \label{app3}

In this Appendix, we show the details of the kite topology discussed in Section~\ref{sec:fac}. The relevant polynomials are given by (with $z_1=z_2=z_3=0$)
\begin{align}
\mG(z_4,z_5) &\equiv 4\nG(l_1,l_2,p) = - 2m^6 + m^4 (s + z_4 + z_5) \nn\\
&+ m^2 (2 z_4 z_5 - s z_4 - s z_5) + z_4 z_5 (s-z_4-z_5) \,, \nn
\\
\mG_1(z_5) &\equiv -4\nG(l_1,p) = (z_5-s)^2+m^4-2m^2(z_5+s) \,,
\end{align}
and the $u$-function is
\begin{equation}
    u(z_4,z_5) = z_4^{\delta_1} z_5^{\delta_2} \left[ \mG(z_4,z_5) \right]^{-\epsilon} .
\end{equation}
To reveal the singularities at $\infty$, we employ the variable changes $z_4=1/t_4$ and $z_5=1/t_5$. The resulting polynomials are
\begin{align}
\mG_{\infty \infty } &\equiv t_4^2t_5^2\  \mG\left(1/t_4,1/t_5\right) = (-2 m^6 +m^4 s )t_4^2 t_5^2 \nn\\ & 
+(m^4 -m^2 s) t_4 t_5^2 + (m^4 -m^2 s) t_4^2 t_5+2 m^2 t_4 t_5 \nn\\ &
+s t_4 t_5-t_4-t_5 \,,\nn\\
\mG_{\infty 0} &\equiv t_4^2\  \mG\left(1/t_4,z_5\right)=-2 m^6 t_4^2 +m^4 s t_4^2+m^4 t_4 \nn\\ & 
+m^4 t_4^2 z_5-m^2 s t_4-m^2 s t_4^2 z_5+2 m^2 t_4 z_5+s t_4 z_5 \nn\\ &
-t_4 z_5^2-z_5 \,,\nn\\
\mG_{0\infty } &\equiv t_5^2\  \mG\left(z_4,1/t_5\right) =\mG_{\infty 0}(t_4 \to t_5, z_5 \to z_4) \,.
\label{eq:Ginf}
\end{align}

The four master integrals can be expressed as $\nd \log$-forms
\begin{align}
	&\varphi_{1}=\nd \log(z_4)\wedge\nd \log(z_5) \,, \nn\\
	&\varphi_{2}=\nd \log(\tau[z_4,m^2;r_{1;\pm}])\wedge\nd \log\left(\frac{z_5-r_{5+}}{z_5-r_{5-}}\right) , \nn\\
	&\varphi_{3}=-\nd \log(\tau[z_4,\infty;r_{1;\pm}])\wedge\nd \log\left(\frac{z_5-r_{5+}}{z_5-r_{5-}}\right) , \nn\\
	&\varphi_{4}=-\nd \log(\tau[z_5,\infty;r_{1;\pm}])\wedge\nd \log\left(\frac{z_4-r_{4+}}{z_4-r_{4-}}\right),  \label{eq:dlog}
\end{align}
where the various roots of quadratic polynomials are given by
\begin{align}  
	&r_{1;\pm}\equiv  r_{\pm}[\mG_1;z_5] \,, \quad r_{4\pm}(z_5)\equiv r_{\pm}[\mG;z_4]\,, \quad r_{5\pm}(z_4)\equiv r_{\pm}[\mG;z_5]\,, \nn\\
    &r_{5+}(\infty)=\infty \,, \quad r_{5-}(\infty)=0 \,, \quad r_{5\pm}(m^2)=m^2 \,.\label{eq:polevalue}
\end{align}
Note that $\varphi_3$ and $\varphi_4$ are related by an exchange symmetry under $z_4 \leftrightarrow z_5$, that we will employ later.

The poles and the relevant variables after factorization are given as
\begin{align}
 &(0,0): \bm{x}^{(1)} \,, \quad (m^2,m^2): \bm{x}^{(2,3)} \,, \quad (\infty,0): \bm{x}^{(4,5,6)} \,, \nn\\
 &(0,\infty): \bm{x}^{(7,8,9)} \,, \quad (\infty,\infty): \bm{x}^{(10,11,12)} \,.
\end{align}
The factorization transformations are related to the following grouping of the 3 factors in the $u$-function:
\begin{align}
\bm{x}^{(2)} &: (\{z_4-m^2,z_5-r_{5+}\} , \{z_5-r_{5-}\}),\nn\\
\bm{x}^{(3)} &: (\{z_4-m^2,z_5-r_{5-}\} , \{z_5-r_{5+}\}),\nn\\
\bm{x}^{(4)} &: (\{t_4\} , \{z_5,z_5-r_-[\mG_{\infty0};z_5]\}),\nn\\
\bm{x}^{(5)} &: (\{t_4,z_5-r_-[\mG_{\infty0};z_5]\},\{z_5\}),\nn\\
\bm{x}^{(6)} &: (\{z_5-r_-[\mG_{\infty0};z_5] \}, \{t_4,z_5\}),\nn\\
\bm{x}^{(7)} &: (\{t_5\} , \{z_4,z_4-r_-[\mG_{0\infty};z_4]\}),\nn\\
\bm{x}^{(8)} &: (\{t_5,z_4-r_-[\mG_{0\infty};z_4]\}\{z_4\} ),\nn\\
\bm{x}^{(9)} &: (\{z_4-r_-[\mG_{0\infty};z_4]\} , \{z_4,t_5\}),\nn\\
\bm{x}^{(10)} &: (\{t_4\} , \{t_5,t_4-r_+[\mG_{\infty\infty};t_4]\}),\nn\\
\bm{x}^{(11)} &: ( \{t_4,t_4-r_+[\mG_{\infty\infty};t_4]\},\{t_5\} ),\nn\\
\bm{x}^{(12)} &: (\{t_4-r_+[\mG_{\infty\infty};t_4]\} , \{t_4,t_5\}).
\end{align}
The explicit transformations for the pole $(m^2,m^2)$ are:
\begin{align}
	z_4 &\to x_1^{(2)} x_2^{(2)} +r_{+}[\mG(x_1^{(2)},x_2^{(2)});x_1^{(2)}] \,, \quad z_5 \to x_2^{(2)} \,,\nn\\
	z_5 &\to x_1^{(3)} x_2^{(3)} +r_{+}[\mG(x_1^{(3)},x_2^{(3)});x_2^{(3)}] \,, \quad z_4 \to x_2^{(3)} \,.
\end{align}
For the pole $(\infty,0)$:
\begin{align}
t_4 &\to x_1^{(4)} x_2^{(4)} \,, \quad z_5 \to x_2^{(4)} \,, \nn
\\
t_4 &\to x_1^{(5)} \,, \quad z_5 \to x_1^{(5)} x_2^{(5)} \,, \nn
\\
t_4 &\to x_1^{(6)} x_2^{(6)} + r_{+}[\mG_{\infty0}(x_1^{(6)},x_2^{(6)});x_1^{(6)}] \,, \quad z_5\to  x_2^{(6)} \,.
\end{align}
The factorization transformations of the pole $(0,\infty)$ is similar as the above due to the $z_4 \leftrightarrow z_5$ symmetry, and we do not show them explicitly. Finally, for the pole $(\infty,\infty)$, we have
\begin{align}
t_4 &\to x_1^{(10)} x_2^{(10)} \,, \quad t_5\to x_2^{(10)} \,,
\\
t_4 &\to  x_1^{(11)} \,, \quad t_5\to x_1^{(11)} x_2^{(11)} \,,
\\
t_4 &\to x_1^{(12)} x_2^{(12)} + r_{+}[\mG_{\infty\infty}(x_1^{(12)},x_2^{(12)});x_1^{(12)}] \,, \quad t_5\to x_2^{(12)} \,.
\end{align}

Note that in all the above transformations, we have shifted the pole to $\bm{\rho}^{(\alpha)}=(0,0)$. Namely, the $u$-function can be written as
\begin{equation}
    u(\bm{x}^{(\alpha)}) = \bar{u}_\alpha(\bm{x}^{(\alpha)}) \left( x_1^{(\alpha)} \right)^{\gamma_1^{(\alpha)}} \left( x_2^{(\alpha)} \right)^{\gamma_2^{(\alpha)}} \,.
\end{equation}
The corresponding $u$-powers are given by (recall that $\bm{\gamma}^{(\alpha)} = \gamma_1^{(\alpha)} \gamma_2^{(\alpha)}$
\begin{align}
&\bm{\gamma}^{(1)}=\delta_1 \delta_2 \,, \quad \bm{\gamma}^{(2,3)}=(-2\ep)(-\ep) \,, \quad \bm{\gamma}^{(7,8,9)}= \bm{\gamma}^{(4,5,6)}\Big|_{\delta_1 \leftrightarrow \delta_2} \,,\nn\\
&\bm{\gamma}^{(4)}= \left(\ep-\delta _1+\delta _2\right)\left(2 \ep -\delta_1\right) , \quad \bm{\gamma}^{(10)}=\left(3\ep-\delta _1-\delta _2\right) \left(2 \ep -\delta_1\right),\nn\\
 &\bm{\gamma}^{(5)}= \left(\ep-\delta _1+\delta _2\right)\delta_2 \,, \quad \bm{\gamma}^{(11)}=\left(3\ep-\delta _1-\delta _2\right) \left(2 \ep -\delta_2\right),\nn\\
 &\bm{\gamma}^{(6)}=\left(\ep-\delta _1+\delta _2\right)(-\ep ) \,, \quad \bm{\gamma}^{(12)}=\left(3\ep-\delta _1-\delta _2\right)(-\ep) \,.
\end{align}
The residues $C^{(\bm{-1})}_I$ of $\varphi_{I}$ at each $\bm{\rho}^{(\alpha)}$ are given by
\begin{align}
	&\varphi_{1}:\{1,0,0,-1,1,0,-1,1,0,1,-1,0\}\,,\nn\\
	&\varphi_{2}:\{0,-1,-1,0,0,0,0,0,0,0,0,0\}\,,\nn\\
	&\varphi_{3}:\{0,0,0,1,0,-1,0,0,0,-1,0,1\}\,,\nn\\
	&\varphi_{4}:\{0,0,0,0,0,0,1,0,-1,0,1,-1\}\,.  \label{eq:commonpole2l}
\end{align}
The zero entries mean that the corresponding integrands are not singular at those poles.

The elements of the $\eta$-matrix, $\eta_{IJ} = \braket{\varphi_I | \varphi_J}$ can be easily obtained using Eq.~\eqref{eq:LP}. The inverse matrix is given by
\begin{align}
	\eta^{-1} = \begin{pmatrix}
		\frac{\delta _1 \delta _2 \left(-\delta _1-\delta _2+\epsilon \right)}{\epsilon } & 0 & -2 \delta _1 \delta _2 & -2 \delta _1 \delta _2 \\
		0 & \epsilon ^2 & 0 & 0 \\
		-2 \delta _1 \delta _2 & 0 & -2 \epsilon  \left(\delta _2+\epsilon \right) & -\epsilon  \left(\delta _1+\delta _2+\epsilon \right) \\
		-2 \delta _1 \delta _2 & 0 & -\epsilon  \left(\delta _1+\delta _2+\epsilon \right) & -2 \epsilon  \left(\delta _1+\epsilon \right) \\
	\end{pmatrix} .
\end{align} 
For the symbol letters contained in $(\nds\Omega)_{13} = \braket{\dot{\varphi}_1 | \varphi_J} \big( \eta^{-1} \big)_{J3}$, we need to compute $\braket{\dot{\varphi}_1|\varphi_1}$, $\braket{\dot{\varphi}_1|\varphi_3}$ and $\braket{\dot{\varphi}_1|\varphi_4}$. Due to the exchange symmetry, $\braket{\dot{\varphi}_1|\varphi_4}$ can be obtained from $\braket{\dot{\varphi}_1|\varphi_3}$ by $\delta_1 \leftrightarrow \delta_2$.

According to Eq.~\eqref{eq:commonpole2l}, the term $\braket{\dot{\varphi}_1|\varphi_1}$ receives $n$-SP contributions from the poles $\rho^{(1,5,8)}$. Note that $\rho^{(4,7,10,11)}$ gives $\nd \log C=0$ and does not contribute to the symbol letters.
In the contributing poles, $\rho^{(8)}$ are apparently related to $\rho^{(5)}$ by the exchange symmetry. Therefore, the genuinely independent contributions to $\braket{\dot{\varphi}_1|\varphi_1}$ are that from $\rho^{(1)}$:
\begin{align}
-\frac{\ep}{\delta_1 \delta_2} \left( 2\log(m^2) + \log(s-2m^2) \right) ,
\end{align}
and that from, e.g., $\bm{\rho}^{(5)}$:
\begin{align}
-\frac{\ep}{(\ep-\delta_1+\delta_2) \delta_2} \left(\log(m^2) + \log(s-m^2) \right) .
\end{align}
The term $\braket{\dot{\varphi}_1|\varphi_3}$ receives $(n-1)$-SP contributions from $\rho^{(4,5,7,8)}$. Again, the only independent non-zero contribution comes from, e.g., $\rho^{(5)}$, and can be written as
\begin{align}
\frac{1}{\ep-\delta_1+\delta_2} \left( \log(m^2) + \log(s-m^2) \right). \label{eq:NPTCLetter}
\end{align}

Combining $\braket{\dot{\varphi}_1|\varphi_J}$ and $\big( \eta^{-1} \big)_{J3}$, we are ready to obtain the result for $(\nds\Omega)_{13}$. For simplicity we take $\delta_1=\delta_2=\delta$. In this case $\bra{\varphi_3} = \bra{\varphi_4}$ and we have only 3 master integrals. The result reads
\begin{align}
(\nds\Omega)_{13} = 4\ep \left( 2\log(s-2m^2) - 3\log(s-m^2) + \log(m^2) \right) .
\end{align}
It is interesting to note that the result does not depend on $\delta$. We have checked this result by comparing it to the differential equations obtained from the traditional IBP, and find agreement. All other elements in $(\nds\Omega)$ can be easily read out since the integrands are 0-form or 1-form after maximal cut.

\end{appendix}


\bibliographystyle{scpma}
\bibliography{intnum_symbol}

\end{multicols}
\end{document}